\documentclass[3p,times,sort&compress]{elsarticle}

\usepackage{ecrc}
\usepackage{amsmath}

\volume{00}

\firstpage{1}

\journalname{Computer Physics Communications}

\runauth{D. Schouten, et al.}

\jid{procs}

\jnltitlelogo{Computer Physics Communications}

\CopyrightLine{2014}{Published by Elsevier Ltd.}

\usepackage{amssymb}

\usepackage[figuresright]{rotating}

\usepackage{slashed}

\newcommand{\m}{\mathcal{M}}
\newcommand{\msqr}{\mathcal{M}^{2}}
\newcommand{\fb}{\ensuremath{\,\textrm{fb}^{-1}}}
\newcommand{\ttH}{\ensuremath{t\bar{t}H}}
\newcommand{\ttHbb}{\ensuremath{t\bar{t}H(\to b\bar{b})}}
\newcommand{\ttbb}{\ensuremath{t\bar{t}b\bar{b}}}
\newcommand{\ttcc}{\ensuremath{t\bar{t}c\bar{c}}}
\newcommand{\ttjj}{\ensuremath{t\bar{t}jj}}
\newcommand{\sy}{\ensuremath{{\bf Y}}}
\newcommand{\sx}{\ensuremath{{\bf X}}}
\newcommand{\vy}{\ensuremath{\vec{y}}}
\newcommand{\vx}{\ensuremath{\vec{x}}}
\newcommand{\met}{\ensuremath{\slashed{E}_T}}

\begin{document}

\begin{frontmatter}



\dochead{}

\title{Accelerated Matrix Element Method with Parallel Computing}


\author[1]{D. Schouten}
\author[2]{A. DeAbreu}
\author[2]{B. Stelzer}
\address[1]{TRIUMF, 4004 Wesbrook Mall, Vancouver, BC}
\address[2]{Department of Physics, Simon Fraser University, 8888 University Dr, Burnaby, BC}

\begin{abstract}
The matrix element method utilizes {\it ab initio} calculations of
probability densities as powerful discriminants for 
processes of interest in experimental particle physics. The method has already
been used successfully at previous and current collider
experiments. However, the computational complexity of this method for
final states with many particles and degrees of freedom sets it at a
disadvantage compared to supervised classification methods such as
decision trees, $k$ nearest-neighbour, or neural  
networks. This note presents a concrete implementation of the matrix
element technique using graphics processing units. Due to the
intrinsic parallelizability of multidimensional integration, dramatic
speedups can be readily achieved, which makes the matrix element
technique viable for general usage at collider experiments. 
\end{abstract}


\end{frontmatter}


\section{Introduction}
\label{sec:introduction}

The matrix element method (MEM) in experimental particle physics is a unique
analysis technique for characterizing collision events. When used to define
a discriminant for event classification, it differs
from supervised multivariate methods such as neural networks, decision trees,
$k$-NN, and support vector machines in that it employs unsupervised {\it ab initio}
calculations of the probability density $P_{i}$ that an observed
collision event with a particular final state arises from $2\to N$
scattering process $i$. Furthermore, the strong connection of this technique to the
underlying particle physics theory provides key benefits compared to more
generic methods:
\begin{enumerate}
  \item the probability density $P_{i}$ directly depends on the physical parameters of
    interest;
  \item it provides a most powerful test statistic for discriminating between
    alternative hypotheses, namely $P_i/P_j$ for hypotheses $i$ and $j$, by the Neyman-Pearson lemma;
  \item it avoids tuning on unphysical parameters for analysis optimization\footnote{Rather, optimization is determined by
    theoretical physics considerations, such as inclusion of higher
    order terms in the matrix element, or improved modeling of
    detector resolution.};
  \item it requires no training, thereby mitigating dependence on large samples of simulated events.
\end{enumerate}
The MEM was first studied in \cite{Konda} and was heavily utilized by
experiments at the Tevatron for $W$ helicity \cite{d0:whelicity} and top mass
\cite{cdf:topmass,d0:topmass} measurements, and in the observation of single top
production \cite{cdf:st}, for example. It has also been used in Higgs
searches at the Tevatron \cite{cdf:higgs} and at the Large Hadron Collider (LHC) by both CMS
\cite{cms:higgs} and ATLAS \cite{atlas:higgsww} collaborations. Good
introductions to the MEM in the context of top mass measurements can
be found in \cite{fiedler, grohsjean}. The MEM has also been extended in 
a general framework known as {\sc MadWeight} \cite{mw}. 

The MEM derives its name from the evaluation of $P_{i}$:
\begin{equation}
  \label{eq:mem_def}
  P_{i} = \frac{1}{\sigma_{i}} \displaystyle\sum_{\textrm{flavor}} \int_{V_{n}} \msqr_{i}(\sy)\,
  \frac{f_{1}(x_{1},Q^{2})\,f_{2}(x_{2},Q^{2})}{|\vec{q}_{1}|\cdot|\vec{q}_{2}|}\, d\Phi_n(q_1+q_2;y_1,..,y_n),
\end{equation}
where $\m_{i}$ is the the Lorentz invariant matrix element for the $2\to
n$ process $i$, $\sy$ is shorthand notation for all the momenta $\vy$ of
each of the $n$ initial and final state particles, $f_{1}$ and
$f_{2}$ are the parton distribution functions (PDF's) for the
colliding partons, $\sigma$ is the overall normalization
(cross-section), and
\begin{equation}
d\Phi_n(q_1+q_2;y_1,..,y_n)=(2\pi)^4\delta^4(q_1+q_2-\sum_{i=1}^n y_i)\prod_{i=1}^n \frac{d^3y_i}{(2\pi)^32E_i}
\end{equation}
is the n-body phase space term. The momenta of the colliding partons are
given by $q_{1}$ and $q_{2}$, and the fractions of the proton beam
energy are $x_{1}$ and $x_{2}$, respectively. The sum 
in Equation  (\ref{eq:mem_def}) indicates a sum over all relevant
flavor combinations for the colliding partons. 

The association of the partonic momenta \sy\ with the measured
momenta \sx\ is given by a transfer function (TF), $T(\vx;\vy)$ for each
final state particle. The TF provides the conditional probability
density function for measuring \vx\ given parton momentum \vy. Thus,
\begin{equation}
  \hat{p}_{i} = \int P_{i}\, T(\sx;\sy)\; d\sy, 
\end{equation}
is the MEM probability density for an observed event to arise from process $i$
assuming the parton $\to$ observable evolution provided by the TF's.
For well-measured objects like photons, muons and electrons, the TF is
typically taken to be a $\delta$-function. For unobserved particles
such as neutrinos, the TF is a uniform distribution. The TF for jet
energies is often modeled with a double Gaussian function, which
accounts for detector response (Gaussian core) and also for parton
fragmentation outside of the jet definition (non-Gaussian tail). To
reduce the number of integration dimensions, the jet directions are
assumed to be well-modeled so that
$T(\theta_{x},\phi_x;\theta_{y},\phi_y) =
\delta(\theta^x-\theta^y)\delta(\phi^x-\phi^y)$.  

Despite the advantages provided by the MEM enumerated above, an important
obstacle to overcome is the computational overhead in evaluating $\geq 1$
multi-dimensional integrals for each collision event. For complex
final states with many degrees of freedom (eg., many particles with
broad measurement resolution, or unobserved particles), the time needed to
evaluate $\hat{p}_{i}$ can exceed many minutes.  In realistic use cases, the
calculations must be performed multiple times for each event, such as
in the context of a parameter estimation analysis where $\hat{p}_{i}$ is
maximized with respect to a parameter of interest, or for 
samples of simulated events used to study systematic biases with varied
detector calibrations or theoretical parameters. 
For large samples of events, the computing time can be prohibitive,
even with access to powerful computer clusters\footnote{As an example,
consider an MEM-based analysis of $\ttHbb$
using 300\fb\ of data at LHC during Run II. The total estimated 
sample size after a simple dilepton + $b$-jet final state selection, including the
irredubicible $\ttbb$ background, is 30$k$ events. Assuming O(5)
minutes to evaluate both $\hat{p}_{\ttbb}$ and $\hat{p}_{\ttHbb}$ for
each event, this implies 2.5$k$ CPU hours needed
for the MEM, for just one pass through the collected data 
sample. It is reasonable to assume a factor of O(50) in CPU time required to
study all systematic uncertainties with Monte Carlo simulations.}. 
Therefore, overcoming the computation hurdle is a relevant goal. 

This paper presents an implementation of the MEM using graphics
processing units (GPU's). The notion of using GPU's for evaluating
matrix elements in a multidimensional phase space has been
investigated previously \cite{heget}, although not in the context of
the MEM. In order to ascertain the improvements in
computing time when utilizing GPU's, the MEM was applied in the context
of a simplified $t\bar{t}H(\to b\bar{b})$ search in LHC Run
II. Studying the $t\bar{t}H$ process is important in its own
right \cite{brout,higgs,hagen,yr:ii,degrande}. Due to the complexity of
the final state for this process, it is also an interesting use case
in which to study the feasibility of the MEM with the improvements from
highly parellelized multi-dimensional integration. 

The note is organized as follows: in Section \ref{sec:GPU}, the
applicability of GPU architectures to the computational 
problem at hand is briefly outlined. In Section \ref{sec:ttH} a simplified
$\ttH$ analysis is presented, which will be used to benchmark the
improved computational performance afforded by modern GPU's. In Section
\ref{sec:results} the specific implementation is outlined together
with a summary of the results obtained from a number of GPU and CPU
architectures. Further details of the codes are listed in \ref{app:code}.  

\section{Parallelized Integrand Evaluation}
\label{sec:GPU}

For dimensions $\geq$ 3, evaluation of multidimensional integrals is typically
only feasible using Monte Carlo methods. In these methods, the integrand
\begin{equation}
  I = \int_{V_{m}} f(\vx)\; d\vx
\end{equation}
is approximated by a sum over randomly sampled points in the
$m$-dimensional integration volume $V_{m}$
\begin{equation}
  \label{eq:sum}
S_{N} \equiv V_m\, \underbrace{\frac{1}{N} \displaystyle
  \sum_{i=1}^{N} f(\vx_{i})}_{\equiv\; \overline{f}}, 
\end{equation}
which converges to $I$ by the law of large numbers. The residual error after
evaluating $N$ points is determined by
\begin{equation}
  \Delta S_{N} \approx \frac{V_m}{\sqrt{N}}\underbrace{\left(\frac{1}{N-1} \displaystyle \sum_{i=1}^{N}
  (f(\vx_i) - \overline{f})^{2}\right)}_{\equiv\; \sigma_{f}}.
\end{equation}
This error estimate is not a strict upper bound, and there can be
significant departures from it depending on the function $f(\vx)$. Modifications
to the simplest Monte Carlo sampling employ stratified and 
importance sampling techniques to improve this error estimate, by
ensuring that regions in which the function varies greatly are
sampled more frequently. One such approach is given by the {\sc Vegas}
algorithm \cite{lepage:1978,lepage:1980}. For all such Monte Carlo
integration algorithms, there is a trivial parallelization that can be
achieved by evaluating the integrand $f(x_i)$ at points
$\{x_{i}\}_{i=1,..,N}$ simultaneously, since the evaluation of the
integrand at each point $x_i$ is independent of all other points
$\{x_j\}_{j\neq i}$. 

This mode of parallel evaluation is known as data parallelism, which
is achieved by concurrently applying the same set of instructions to
each data element. In practice, evaluating the functions
used in the MEM involves conditional branching, so that the
integrand calculation at each $x_i$ does not follow an identical
control flow. Nevertheless, it is instructive to proceed with the
ansatz of strict data parallelism. 

Data parallelism maps very well to the single instruction, multiple
data (SIMD) architecture of graphics processing units (GPU's). Modern
GPU's contain many individual compute units organized in thread
units. Within each thread unit, all threads follow the same
instruction sequence\footnote{This has
implications for code with complicated control flow, since threads
will be locked waiting for other threads in the same unit to be
syncronized in the instruction sequence. Careful tuning of the MEM
function control flow and the thread unit sizes may improve the
performance.}, and have access to a small shared memory cache in addition to
the global GPU memory. 

The advent of general purpose programming on GPU's
(GPGPU) has vastly increased the computing capability
available on a single workstation, especially for data parallel
calculations such as in the MEM. Two languages have gained traction
for GPGPU, namely CUDA \cite{cuda} (restricted to GPU's manufactured by NVidia)
and OpenCL \cite{opencl}. Both languages are based on
C/C++\footnote{In this work, the AMD Static C++ extensions to
  OpenCL \cite{amd:2012} are used.}.

\section{$\ttHbb$ Search}
\label{sec:ttH}

\begin{figure}[b!]
\centering
  \begin{tabular}{@{}c@{}}{\includegraphics[width=0.3\textwidth]{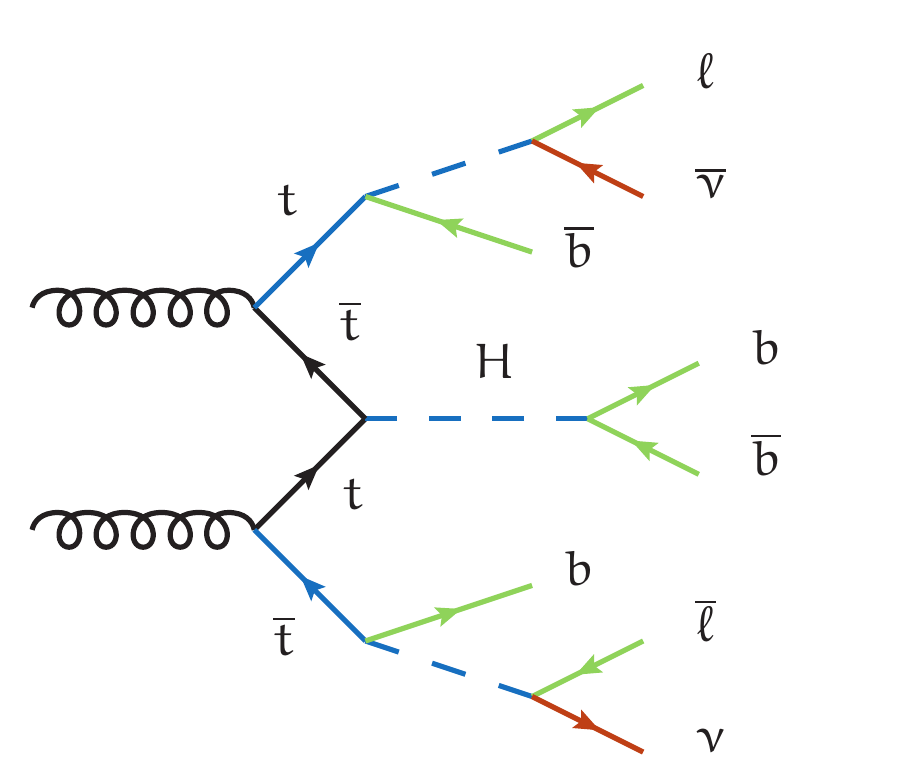}}\end{tabular}
  \begin{tabular}{@{}c@{}}{\hspace*{2cm}\includegraphics[width=0.3\textwidth]{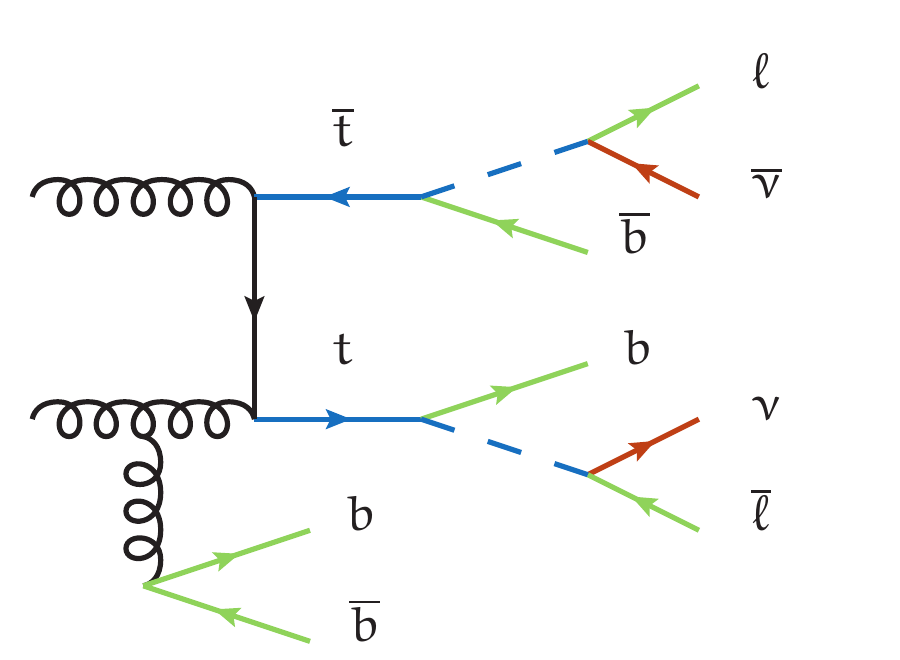}}\end{tabular}

\caption{\label{fig:feynman} Representative Feynman diagrams for
  \ttHbb\ production (left) and the irreducible \ttbb\ background
  (right).}
\end{figure}

The fact that the Higgs coupling to the top quark is $\approx$ 1
hints at a special role played by the top quark in electroweak
symmetry breaking.  Analysis of \ttHbb\ production at the LHC can
provide a powerful direct constraint on the fermionic (specifically, the top)
couplings of the Higgs boson, with minimal model-dependence. The
dominant backgrounds to this process, assuming at least one leptonic top decay,
arise from the irreducible \ttbb\ background as well as the \ttcc\ and
\ttjj\ backgrounds via `fake' $b$-tagged jets. The association of
observed jets to the external lines of the leading order (LO) Feynman
diagrams in Figure \ref{fig:feynman} also gives rise to a combinatoric
dilution of the signal, since there is an increased probability that a
random pair of $b$ partons in a \ttbb\ event will have $m_{bb\,'} \approx m_H$. For
the fully hadronic $t\bar{t}$ decay, there are 4! $\times$ 4! / 2=
288 combinations, assuming fully efficient $b$-tagging. This benchmark
study is restricted to the dileptonic $t\bar{t}$
decay mode, to reduce the combinatoric and also the large
$W$+jet(s) and QCD backgrounds.  

For the dilepton channel, the observable signature of the final state  is
$b\bar{b}\ell\bar{\ell}$ + $\met$, where $\met = \vec{p}_{T}^{\nu} +
\vec{p}_{T}^{\bar{\nu}}$. It is not possible to constrain the $z$
components of the neutrino momenta, and using transverse momentum
balance removes only two of the remaining four degrees of freedom from
the $x$ and $y$ components. Due to the broad resolution of the
measured jet energy, there are also four degrees of freedom for the
energy of the four $b$ quarks in the final state, so that the MEM
evaluation implies an 8-dimensional integration:
\begin{eqnarray}
  \label{eq:integrand}
  \hat{p}_{i} & = & \frac{1}{\sigma_{i}} \displaystyle\sum_{\textrm{jet comb.}} \displaystyle\sum_{\textrm{flavor}} \int \msqr_{i}(\sy)\,
  \frac{f_{1}(x_{1},Q^{2})\,f_{2}(x_{2},Q^{2})}{|\vec{q}_{1}|\cdot|\vec{q}_{2}|}\; \Phi \cdot
  \\ \nonumber & & \; \; \; \;
  \delta\left(p_x^\nu - \met^x -
  p_x^{\bar{\nu}}\right)\,\delta\left(p_y^\nu - \met^y -
  p_y^{\bar{\nu}}\right)\, d^3\vec{p}_\nu\,d^3\vec{p}_{\bar{\nu}}
  \prod_{j=1}^{N_{\textrm{jet}}=4}
  \textrm{T}\,(E^{\textrm{jet}}_{j};E_{j})\cdot (E_j^2\,
  \textrm{sin}\,\theta_{j})\, dE_{j},
\end{eqnarray}
where $\Phi = (2\pi)^4\delta^4(q_1+q_2-\sum_{i=1}^n p_y^i)\prod_{i=1}^n
\frac{1}{(2\pi)^32E_i}$, and the integrals over the lepton momenta are
removed by assuming infinitesimal measurement resolution. The outer
sum is over all permutations of assigning measured jets to partons
in the matrix element. A transformation to spherical coordinates
has been performed $d^{3}\vec{p} \to p^{2}\,
\textrm{sin}(\theta)\, dp\, d\theta\, d\phi$ for the jets, and $E$ is
set to $|p|$.

The behaviour of the matrix element function $\m({\bf Y})$ is strongly influenced by
whether or not the internal propagators are on shell. It is 
difficult for numerical integration algorithms to efficiently map out
the locations in momentum space of the external lines for which the
internal lines are on shell. Therefore, it is advantageous to transform
integration over the neutrino momenta to integrals over $q^{2}$ (where
$q$ is the four momentum) of the top quark and $W$ boson
propagators, so that the poles in the integration volume are along
simple hyperplanes. This leads to the following coupled equations:
\begin{eqnarray}
\label{eq:dilepsolv}
\slashed E_{x} &= &p^{\nu}_{x} + p^{\bar{\nu}}_{x} \nonumber \\
\slashed E_{y} &= &p^{\nu}_{y} + p^{\bar{\nu}}_{y} \nonumber \\
q_{W^{+}}^{2} & = &(E_{\ell^{+}} + E_{\nu})^{2} - (p^{\ell^{+}}_{x} +
p^{\nu}_{x})^{2} - (p^{\ell^{+}}_{y} + p^{\nu}_{y})^{2} - (p^{\ell^{+}}_{z} + p^{\nu}_{z})^{2} \nonumber \\
q_{W^{-}}^{2} & = &(E_{\ell^{-}} + E_{\bar{\nu}})^{2} - (p^{\ell^{-}}_{x} +
p^{\bar{\nu}_{x}})^{2} - (p^{\ell^{-}}_{y} + p^{\bar{\nu}_{y}})^{2} - (p^{\ell^{-}}_{z} + p^{\bar{\nu}_{z}})^{2} \\
q_{t}^{2} & = &(E_{b} + E_{\ell^{+}} + E_{\nu})^{2} - (p^{b}_{x} +
p^{\ell^{+}}_{x} + p^{\nu}_{x})^{2} - \nonumber \\
& & (p^{b}_{y} + p^{\ell^{+}}_{y} + p^{\nu}_{y})^{2} - (p^{b}_{z} +
p^{\ell^{+}}_{z} + p^{\nu}_{z})^{2}  \nonumber \\
q_{\bar{t}}^{2} & = &(E_{\bar{b}} + E_{\ell^{-}} + E_{\bar{\nu}})^{2} -
(p^{\bar{b}}_{x} + p^{\ell^{-}}_{x} + p^{\bar{\nu}}_{x})^{2} - \nonumber \\
& & (p^{\bar{b}}_{y} + p^{\ell^{-}}_{y} + p^{\bar{\nu}}_{y})^{2} - (p^{\bar{b}}_{z} + p^{\ell^{-}}_{z} + p^{\bar{\nu}}_{z})^{2}, \nonumber
\end{eqnarray}
which have been solved analytically in \cite{sonnenschein}. For the
$\ttHbb$ process, there is an additional very narrow resonance from
the Higgs propagator. For the same reasoning as above, the following
transformation of variables for $E_{1}$ and $E_{2}$ are employed,
which are the energies of the $b$-quarks from the Higgs decay, respectively:
\begin{eqnarray}
  \label{eq:mhsolve}
  f &=& (E_{1} + E_{2}) \nonumber \\
  m_{H}^{2} &=& (E_{1} + E_{2})^2 - |\vec{p}_{1}|^{2} - |\vec{p}_{2}|^{2} -
  2\, |\vec{p}_{1}|\, |\vec{p}_{2}|\, \textrm{cos}\,\Delta\theta_{1,2},
\end{eqnarray}
where $|\vec{p}| = \sqrt{E^{2} - m^{2}}$.
Figure \ref{fig:feynman} highlights the internal lines which are used
in the integration. 

\section{Analysis and Results}
\label{sec:results}

\begin{figure}
\centering
\includegraphics[width=0.4\textwidth]{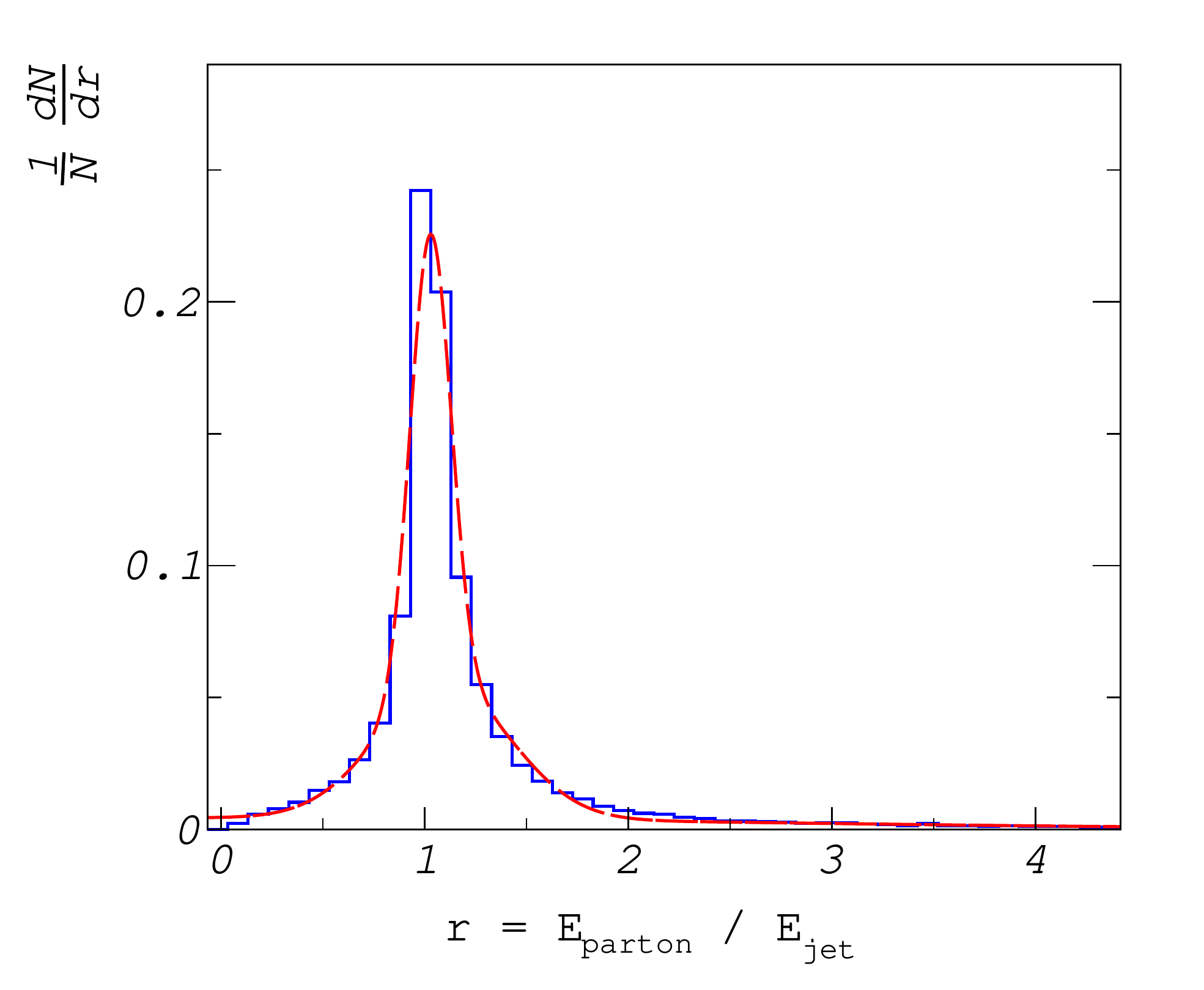}
\caption{\label{fig:tf_jet} The ratio of parton energy to jet energy for
  jets reconstructed with the anti-$k_T$ algorithm and matched to
  partons within $\Delta R < 0.4$. The fitted transfer function is
  also shown.}
\end{figure}

The evaluation of the integrand in Equation  (\ref{eq:integrand}) is broken into
components for the  matrix element $\m({\bf Y})$, the PDF's, the TF's and the
phase space factor. Each of these components is evaluated within a
single GPU ``kernel'' program for each phase space point. Code for
evaluating $\m$ is generated using a plugin developed for {\sc
  MadGraph} \cite{madgraph}. This plugin allows one to export code for an arbitrary $2\to
N$ process from {\sc MadGraph} to a format compatible with OpenCL, CUDA, and standard
C++. This code is based on HELAS functions
\cite{helas,aloha}. Compilation for the various platforms is 
controlled with precompiler flags. Model parameters, PDF grids and
phase space coordinates are loaded in memory and transferred to the
device\footnote{In the case of CPU-only computation, the transfer
step is unnecessary.} (GPU) whereafter the kernel is
executed. The PDF's are evaluated within the kernel using wrapper
code that interfaces with LHAPDF \cite{lhapdf} and 
with the CTEQ \cite{ct10} standalone PDF library. The PDF data is
queried from the external library and stored in $(x,Q^{2})$ grids for
each parton flavor ($d,u,s,c,b$), which are passed to the kernel program. The PDF for an
arbitrary point is evaluted using bilinear
interpolation within the kernel. The precision of the interpolation is
within 1\% of the values directly queried from the PDF library. An
event discriminant $D$ is constructed as 
\begin{equation}
  D = \textrm{log}_{10} \left( \frac{\hat{p}_{\ttH}}{\hat{p}_{\ttbb}}
  \right) 
\end{equation}
and evaluated for a sample of signal (\ttH) and background (\ttbb) events
generated in {\sc MadGraph} and interfaced with {\sc
  Pythia} for the parton shower, \cite{pythia} using the so-called Perugia tune
\cite{perugia}.  Jets are reconstructed using the anti-$k_T$ algorithm
described in \cite{fastjet} with width parameter $d=0.4$. Any jets overlapping
with leptons within $d$ are vetoed, and $b$-tagging is performed by
matching jets to the highest energy parton within $\Delta R =
\sqrt{\Delta\eta^2 + \Delta\phi^2} < d$. A transfer function is
defined for $b$-jets by fitting the ratio of jet energy to the energy of
the matched parton using a double Gaussian distribution, as shown in
Figure \ref{fig:tf_jet}. 

The analysis is performed at two levels, namely
\begin{enumerate}
  \item parton level: using the parton momenta from {\sc
    MadGraph}-generated events directly (by assuming $\delta$-function
    TF's for all final state particles), and averaging over all
    permutations for the assignment of the $b$ partons in each event;
  \item hadron level: using the outputs from {\sc Pythia} and selecting
    events with four $b$-jets, averaging over all the permutations and
    integrating over the full 8-dimensional phase space as in Equation  (\ref{eq:integrand}).
\end{enumerate}

The convenient PyOpenCL and PyCUDA \cite{pycuda} packages are used to
setup and launch OpenCL and CUDA kernels. Using OpenCL one 
can also compile the MEM source for a CPU target, and is thereby able to
parallelize the MEM across multiple cores (see Table \ref{tab:hw_configs}). In order to perform the
numerical integration, a modified {\sc Vegas} implementation in Cython/Python
\cite{lepage:2014} is used. This implementation has a number of improvements
compared to previous versions and, importantly, interfaces with the
provided integrand function by passing the full grid of phase space points as a single
function argument. This allows one to pass the whole integration grid to the
OpenCL or CUDA device at once, which facilitates the desired high degree of
parallelism. The parton and hadron level MEM is performed with various
hardware configurations specified in Table \ref{tab:hw_configs}. In
all configurations except the one labelled $\textrm{GPU}_{x}$, the MEM code used is
identical. For the $\textrm{GPU}_{x}$ case, minor modifications were made to replace
particular array variables with sets of scalar variables. 

\begin{table}
  \centering
  \resizebox{1.0\textwidth}{!}{\begin{tabular}{l|l|c|c}
    \hline
    \hline
    Configuration      & Details & Peak Power & Cost (USD, 2014) \\
    \hline
    CPU                & Intel Xeon CPU E5-2620 0 @ 2.00GHz (single core) using \texttt{gcc} 4.8.1 &  95W & 400 \\
    CPU (MP)           & Intel Xeon CPU E5-2620 0 @ 2.00GHz (six cores + hyperthreading) using AMD SDK 2.9 / OpenCL 1.2 & 95W & 400 \\
    \hline
    GPU               & AMD Radeon R9 290X GPU (2,816 c.u.) using AMD SDK 2.9 / OpenCL 1.2 on Intel Xeon CPU E5-2620 & 295W & 450 \\
    $\textrm{GPU}_{x}$              & same configuration as GPU, but with minor code modifications to accomodate GPU architecture &  &  \\
    \hline
    \hline
  \end{tabular}}
  \caption{\label{tab:hw_configs}Details of the hardware configurations used to
  benchmark the MEM for GPU and (multicore) CPU's. The peak power is
  as reported by the manufacturer. The cost is listed in USD for the
  CPU or GPU only. For the GPU configuration, the code was identical
  to that used for the CPU configurations. For the $\textrm{GPU}_{x}$
  configuration, the code was modified to accommodate the specific
  GPU architecture.}  
\end{table}    

\subsection{Parton Level}
\begin{figure}[h]
\centering
\includegraphics[width=0.4\textwidth]{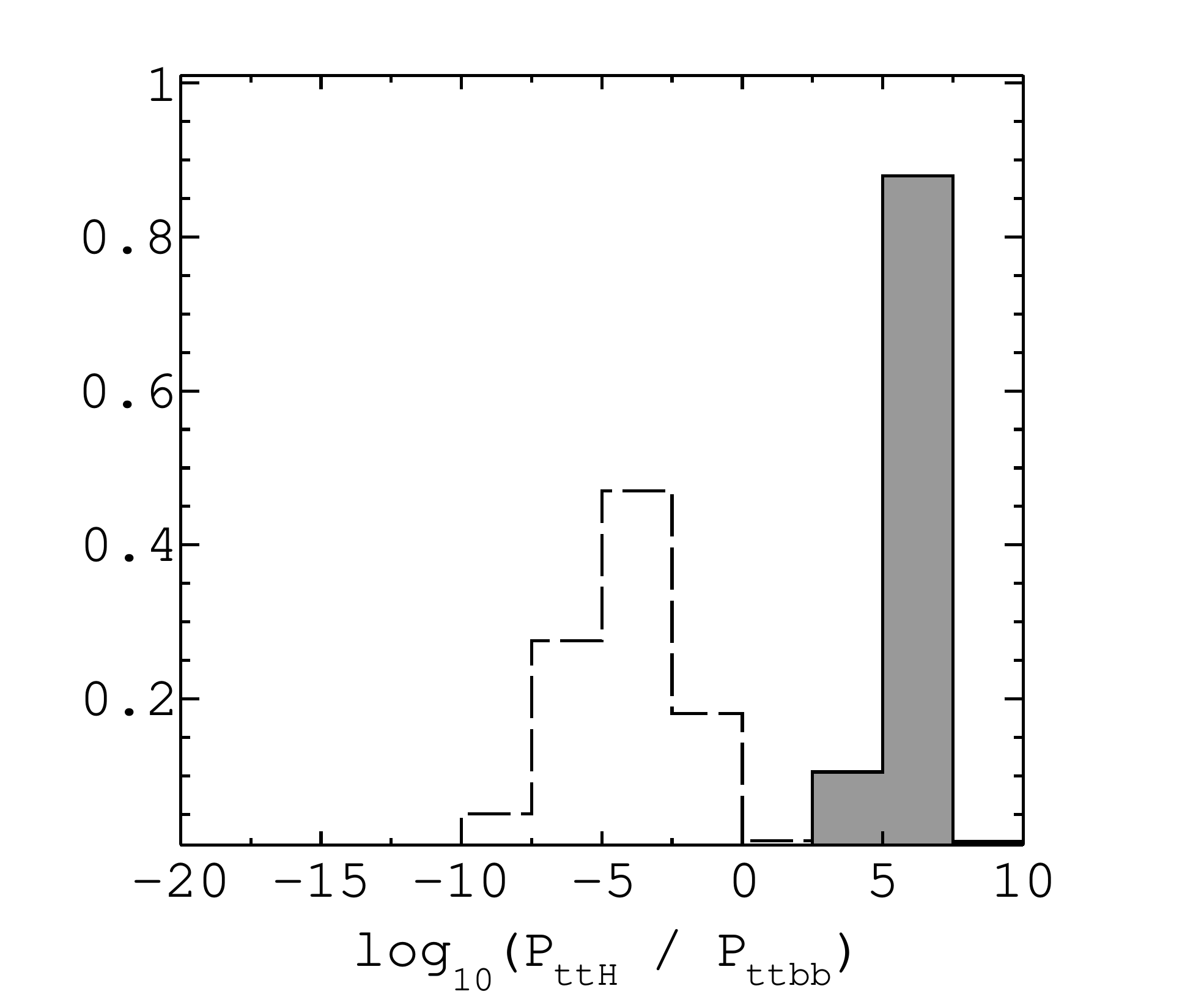}
\includegraphics[width=0.4\textwidth]{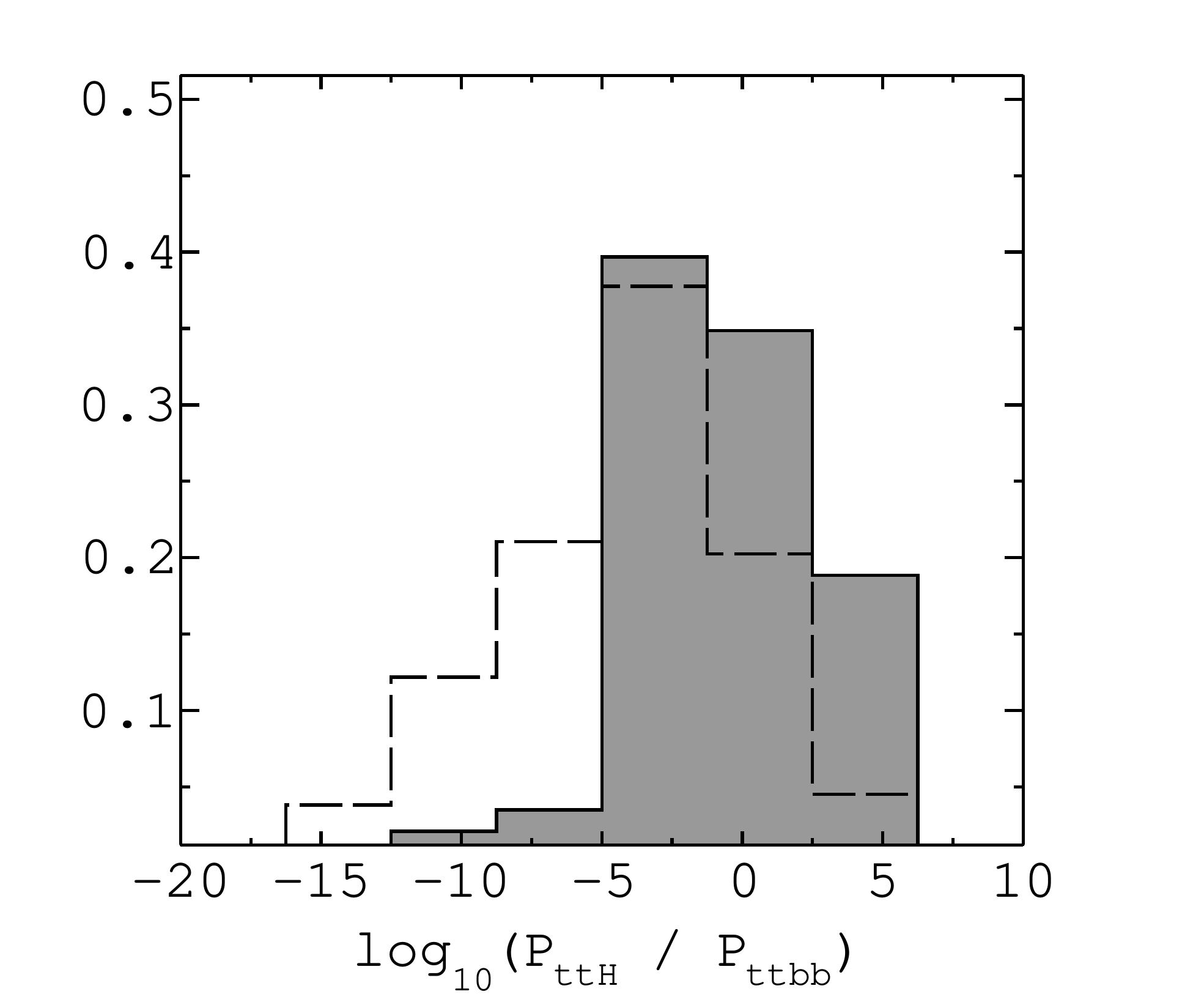} \\ \vspace{-2ex}
\caption{\label{fig:epd}The event discriminant $D$ for
  \ttH\ (filled) and \ttbb\ (dashed line) events at
  parton level (left) and at hadron level (right). The distributions
  are normalized to unit area.}
\end{figure}

Here, the evaluation is performed using the parton momenta, so that
all the transfer functions become $\delta$-functions, and the
evaluation of $D$ does not involve any numberical integration. In this
benchmark, all possible combinations of $b$-quarks in the
final state are summed. The distribution of $D$ for the signal and background
samples is shown in Figure \ref{fig:epd}. A comparison of the
time needed to evaluate $\hat{p}_{i}$ for all events is shown in Table
\ref{tab:time_parton} for various CPU and GPU configurations. 

\begin{table}[h!]
  \centering
  {\footnotesize {\renewcommand{\arraystretch}{1.4} \begin{tabular}{l|c|c|c|c|c}
    \hline \hline
    Process    & CPU  & CPU (MP) & GPU  & $\textrm{GPU}_{x}$ & $\textrm{GPU}_{x}$ / CPU \\
    \hline
    signal     &  255 & 29       &  1.8 & 0.7  &  364  \\
    \hline
    background &  661 & 91       &  12  & 5.4  &  122   \\
    \hline \hline
  \end{tabular}}}
  \caption{\label{tab:time_parton} Processing time, in seconds, required to
    evaluate the matrix elements for 10$^5$ events at parton level, for the various
    configurations detailed in Table \ref{tab:hw_configs}. Using GPU's
    reduces the processing time by a factor greater than 120$\times$ compared
    to a single CPU core for the $\ttbb$ matrix element. }
\end{table}

\subsection{Hadron Level}

The analysis at hadron level is closer to what can be optimally
achieved in a real world collider experiment. Only the momenta of stable,
interacting particles are accessible, and the jet energy resolution (see
Figure \ref{fig:tf_jet}) must be taken into account. The calculation
of $\hat{p}_i$ requires evaluating the eight-dimensional integral in
Equation  (\ref{eq:integrand}). The integration variable transformation for $\ttH$ and
$\ttbb$ matrix element integrals presented in Section
\ref{sec:GPU} are used. At each phase space point in the sum of
Equation  (\ref{eq:sum}), the $\met$ used in Equation (\ref{eq:dilepsolv}) is defined as 
\begin{equation}
  \slashed E_{x,y} = -\left(p^{\ell^{+}}_{x,y} + p^{\ell^{-}}_{x,y} +
  \displaystyle\sum_{j\,\in\,\textrm{jets}} p^{j}_{x,y}\right).
\end{equation}
The processing times per event for the hadron level MEM calculation are
shown in Table \ref{tab:time_hadron}. The relative improvement for the GPU
is significantly reduced compared to the parton level analysis. This
arises from a number of differences for this scenario. First, the
VEGAS stratified sampling and adaptive integration algorithm is run on the
CPU in all cases, which damps the GPU improvements in the integrand
evaluation. Second, in the evaluation of the integral of Equation (\ref{eq:integrand}),
significant additional complexity is demanded to solve Equations
(\ref{eq:dilepsolv}) and (\ref{eq:mhsolve}). Due to the cancellation of
large coefficients in these solutions, double floating point precision
is required, which reduces the GPU advantage since double precision
calculations are performed significantly slower on most
GPU's. Furthermore, the number of intermediate variables is
significantly larger, which is found to increase the number of
processor registers used. Since the number of registers available to each
thread unit (or ``wavefront'' in the parlance of OpenCL) is limited to
at most 256 for the GPU used in this study, 
the overall duty factor of the GPU is significantly reduced, to as low
as 10\%, since the full number of threads available in each block
could not be utilized. It is anticipated that careful tuning of the
code to accommodate GPU architecture could greatly improve the relative
performance.  

\begin{table}[t]
  \centering
  {\footnotesize {\renewcommand{\arraystretch}{1.4} \begin{tabular}{l|c|c|c|c|c}
    \hline \hline
    Process    & CPU & CPU (MP) & GPU & $\textrm{GPU}_{x}$ & $\textrm{GPU}_{x}$ / CPU  \\
    \hline
    signal     &  312 & 36.2 & 7.5 & 5.9 & 52.0 \\
    \hline
    background &  405 & 55.1 & 9.1 & 7.1 & 57.3 \\
    \hline \hline
  \end{tabular}}}
  \caption{\label{tab:time_hadron} Processing time required to
    evaluate the matrix elements for a single event at hadron level, for the various
    configurations detailed in Table \ref{tab:hw_configs}. Note that
    this includes a full 8-dimensional integration over phase space
    for each event. Using GPU's reduces the processing time by at least 50$\times$.}
\end{table}

\section{Conclusions}
\label{sec:conclusions}

The matrix element method can be computationally prohibitive for
certain final states. The benchmark study in this paper has shown that
by exploiting the parallel architectures of modern GPU's,
computation time can be reduced by a factor $\geq 50$ for the
matrix element method, at about 10\% utilization of the GPU. It is anticipated that careful code
modifications can add significant further improvements in speed.  This
can be the subject of future study. However, even with the performance
gains in this benchmark study, it is clear that for the MEM, the 
computing time required with O(10) GPU's is equivalent to a
medium-sized computing cluster with O(400) cores (along with its required
support and facilities infrastructure). This provides the potential to
apply the method generally to searches and measurements with complex
final states in experimental particle physics.

The programs described in this work are generic in nature, such that
GPU-capable MEM code can be readily derived for an arbitrary $2\to N$
process with only few modifications to accommodate transformations of
variables or transfer functions. It is envisaged that future work can
automate the inclusion of NLO matrix elements and transformations of
variables (as in {\sc MadWeight}) for the matrix element method,
thereby providing an optimal methodology for classification and
parameter estimation in particle physics.

\section{Acknowledgements}
\label{sec:acknowledgements}

This research was enabled in part by support provided by WestGrid
(www.westgrid.ca) and Compute Canada / Calcul Canada
(www.computecanada.ca). We acknowledge the support of the Natural
Sciences and Engineering Research Council of Canada (NSERC) and the
Vice President Research Office of Simon Fraser University. 

\clearpage

\appendix

\section{Program Listings}
\label{app:code}

\begin{figure}[h!]
{\scriptsize
\begin{verbatim}
typedef int   a_int_t;
typedef float a_float_t;

#ifdef _CL_CUDA_READY_ ///////////////////////////////////////////////////////// DEVICE
#ifdef _OPENCL_
#define _POW_ pow
#define _SQRT_ sqrt
#define _EXP_ exp
#define _LOG_ log
#define _ATAN_ atan
#define _TAN_ tan
#define _ACOS_ acos
#define _COS_ cos
#define _ASIN_ asin
#define _SIN_ sin
#define _CL_CUDA_HOST_ 
#define _CL_CUDA_DEVICE_ 
#define _CL_CUDA_GLOBAL_ __global 
#define _CL_CUDA_CONSTANT_ __constant
#define _CL_CUDA_KERNEL_ __kernel
#endif

#ifdef _CUDA_
#define _POW_ powf
#define _SQRT_ sqrtf
#define _EXP_ expf
#define _LOG_ logf
#define _ATAN_ atanf
#define _TAN_ tanf
#define _ACOS_ acosf
#define _COS_ cosf
#define _ASIN_ asinf
#define _SIN_ sinf
#define _CL_CUDA_HOST_ __host__
#define _CL_CUDA_DEVICE_ __device__
#define _CL_CUDA_BOTH_ __host__ __device__
#define _CL_CUDA_GLOBAL_
#define _CL_CUDA_CONSTANT_
#define _CL_CUDA_KERNEL_ __global__
#endif
#else //////////////////////////////////////////////////////////////////////////// HOST
#include <cmath>
#define _POW_ std::pow
#define _SQRT_ std::sqrt
#define _EXP_ std::exp
#define _LOG_ std::log
#define _TAN_ std::tan
#define _ATAN_ std::atan
#define _COS_ std::cos
#define _ACOS_ std::acos
#define _SIN_ std::sin
#define _ASIN_ std::asin
#define _CL_CUDA_HOST_ 
#define _CL_CUDA_DEVICE_ 
#define _CL_CUDA_GLOBAL_
#define _CL_CUDA_CONSTANT_
#define _CL_CUDA_KERNEL_
#define _CL_CUDA_IDX_ 0
#endif ////////////////////////////////////////////////////////////////////////////////
\end{verbatim}}
\caption{Listing of common header used to configure the calculations
  for OpenCL, CUDA and C/C++ compilation.}
\label{fig:matcommon.h}
\end{figure}

\begin{figure}
{\scriptsize
\begin{verbatim}
_CL_CUDA_HOST_ _CL_CUDA_DEVICE_ a_float_t pdf(a_int_t fl, 
                                              a_float_t x, 
                                              a_float_t Q,
                                              _CL_CUDA_CONSTANT_ a_float_t pdf_data[], 
                                              _CL_CUDA_CONSTANT_ a_float_t pdf_bounds[])
{
  if(fl < -5 || fl > 5) return 0;
  const a_int_t IOFFSET = (fl - FLAVOR_OFFSET) * (NUM_XSAMPLES_DEF * NUM_QSAMPLES_DEF);
  
  a_float_t pdf_lxmin = pdf_bounds[0];
  a_float_t pdf_lxmax = pdf_bounds[1]; 
  a_float_t pdf_lqmin = pdf_bounds[2]; 
  a_float_t pdf_lqmax = pdf_bounds[3];
  
  x = _LOG_(x);
  Q = _LOG_(Q);
  
  a_float_t dx = (pdf_lxmax - pdf_lxmin) / NUM_XSAMPLES_DEF;
  unsigned ix = 0;
  if(pdf_lxmin < x) ix = static_cast<unsigned int>((x - pdf_lxmin) / dx);
  ix = ix < NUM_XSAMPLES_DEF-1 ? ix : NUM_XSAMPLES_DEF-2;
  
  a_float_t dq = (pdf_lqmax - pdf_lqmin) / NUM_QSAMPLES_DEF;
  unsigned iq = 0;
  if(pdf_lqmin < Q) iq = static_cast<unsigned int>((Q - pdf_lqmin) / dq);
  iq = iq < NUM_QSAMPLES_DEF-1 ? iq : NUM_QSAMPLES_DEF-2;
  
  a_float_t c11x = pdf_lxmin + dx*ix;
  a_float_t c11q = pdf_lqmin + dq*iq;
  
  a_float_t c22x = pdf_lxmin + dx*(ix+1);
  a_float_t c22q = pdf_lqmin + dq*(iq+1);
  
  a_float_t norm = dx*dq;
  
  unsigned int i0j0 = iq*NUM_XSAMPLES_DEF + ix;
  unsigned int i1j0 = (iq+1)*NUM_XSAMPLES_DEF + ix;
  unsigned int i0j1 = iq*NUM_XSAMPLES_DEF + ix + 1;
  unsigned int i1j1 = (iq+1)*NUM_XSAMPLES_DEF + ix + 1;
  
  return (pdf_data[i0j0+IOFFSET] / (norm) * (c22x - x)*(c22q - Q) +
	  pdf_data[i0j1+IOFFSET] / (norm) * (x - c11x)*(c22q - Q) + 
	  pdf_data[i1j0+IOFFSET] / (norm) * (c22x - x)*(Q - c11q) + 
	  pdf_data[i1j1+IOFFSET] / (norm) * (x - c11x)*(Q - c11q));
}
\end{verbatim}}
\caption{Listing of PDF function adapted for OpenCL \& CUDA. The
  function takes as input a uniform grid ($x,Q$) for each parton
  flavor, as provided by querying any third-party PDF set. The PDF at
  any point in the ($x,Q$) plane is then determined using bilinear
  interpolation.}
\label{fig:pdf.cc}
\end{figure}

\begin{figure}
{\scriptsize  
\begin{verbatim}
#ifndef cmplx_h
#define cmplx_h

#include "matcommon.h"

struct a_cmplx_t {
  a_float_t re; a_float_t im;
  
  _CL_CUDA_DEVICE_ a_cmplx_t()                         { }
  _CL_CUDA_DEVICE_ a_cmplx_t(a_float_t x, a_float_t y) { re = x; im = y; }
  _CL_CUDA_DEVICE_ a_cmplx_t(a_float_t x)              { re = x; im = 0; }  
  _CL_CUDA_DEVICE_ a_cmplx_t(const a_cmplx_t& c)       { re = c.re; im = c.im; }
  
  a_cmplx_t& operator=(const a_float_t& x)  { re = x; im = 0; }
  a_cmplx_t& operator=(const a_cmplx_t& c)  { re = c.re; im = c.im; }
};

inline _CL_CUDA_DEVICE_ a_float_t real(a_cmplx_t a) { return a.re; }

inline _CL_CUDA_DEVICE_ a_float_t imag(a_cmplx_t a) { return a.im; }

inline _CL_CUDA_DEVICE_ a_cmplx_t conj(a_cmplx_t a) { return a_cmplx_t(a.re,-a.im); }

inline _CL_CUDA_DEVICE_ a_float_t fabsc(a_cmplx_t a) { return _SQRT_((a.re*a.re)+(a.im*a.im)); }

inline _CL_CUDA_DEVICE_ a_float_t fabsc_sqr(a_cmplx_t a) { return (a.re*a.re)+(a.im*a.im); }

inline _CL_CUDA_DEVICE_ a_cmplx_t operator+(a_cmplx_t a, a_cmplx_t b) { return a_cmplx_t(a.re + b.re, a.im + b.im); }

inline _CL_CUDA_DEVICE_ a_cmplx_t operator+(a_float_t a, a_cmplx_t b) { return a_cmplx_t(a + b.re, b.im); }

inline _CL_CUDA_DEVICE_ a_cmplx_t operator+(a_cmplx_t a) { return a_cmplx_t(+a.re, +a.im); }

inline _CL_CUDA_DEVICE_ a_cmplx_t operator-(a_cmplx_t a, a_cmplx_t b) { return a_cmplx_t(a.re - b.re, a.im - b.im); }

inline _CL_CUDA_DEVICE_ a_cmplx_t operator-(a_float_t a, a_cmplx_t b) { return a_cmplx_t(a - b.re, -b.im); }

inline _CL_CUDA_DEVICE_ a_cmplx_t operator-(a_cmplx_t a) { return a_cmplx_t(-a.re, -a.im); }

inline _CL_CUDA_DEVICE_ a_cmplx_t operator*(a_cmplx_t a, a_cmplx_t b) {
  return a_cmplx_t((a.re * b.re) - (a.im * b.im),
		   (a.re * b.im) + (a.im * b.re));
}

inline _CL_CUDA_DEVICE_ a_cmplx_t operator*(a_cmplx_t a, a_float_t s) { return a_cmplx_t(a.re * s, a.im * s); }

inline _CL_CUDA_DEVICE_ a_cmplx_t operator*(a_float_t s, a_cmplx_t a) { return a_cmplx_t(a.re * s, a.im * s); }

inline _CL_CUDA_DEVICE_ a_cmplx_t operator/(a_cmplx_t a, a_cmplx_t b) {
  a_float_t t=(1./(b.re*b.re+b.im*b.im));
  return a_cmplx_t( ( (a.re * b.re) + (a.im * b.im))*t,
		    (-(a.re * b.im) + (a.im * b.re))*t );
}

inline _CL_CUDA_DEVICE_ a_cmplx_t operator/(a_cmplx_t a, a_float_t s) { return a * (1. / s); }

inline _CL_CUDA_DEVICE_ a_cmplx_t operator/(a_float_t s, a_cmplx_t a) {
  a_float_t inv = s*(1./(a.re*a.re+a.im*a.im));
  return a_cmplx_t(inv*a.re,-inv*a.im);
}

#endif // cmplx_h
\end{verbatim} 
}
\caption{\label{cmplx.h} Listing of complex number type written for
  OpenCL, for which there is no native equivalent.}
\end{figure}



\clearpage

\bibliographystyle{elsarticle-num}
\bibliography{article}







\end{document}